\documentclass[10pt, journal, letterpaper]{IEEEtran}  

\usepackage{graphics} 
\usepackage{epsfig} 
\usepackage{times} 
\usepackage{amsmath} 
\usepackage{mathtools}
\usepackage{amssymb}  
\usepackage{bbold}
\usepackage{color}
\usepackage[ruled,linesnumbered]{algorithm2e} 
\usepackage{subfigure}
\usepackage{pbox}
\usepackage{booktabs}

\newtheorem{proposition}{\textbf{Proposition}}

\newcommand{\rcolor}{black}

\newcommand{\smp}{{\textsc{smpool}}\xspace}

\newcommand{\mmr}{\textit{MR}\xspace} 
\newcommand{\mrm}{\textit{RMG}\xspace} 
\newcommand{\mfr}{\textit{FR}\xspace} 
\newcommand{\mat}{\textit{AT50}\xspace} 
\newcommand{\mpr}{\textit{PRTH}\xspace} 

\title{Modeling the Impact of Network Connectivity on Consensus Security of Proof-of-Work Blockchain}

\author{
\IEEEauthorblockN{
    {Yang Xiao}\IEEEauthorrefmark{1},
      {Ning Zhang}\IEEEauthorrefmark{2},
      {Wenjing Lou}\IEEEauthorrefmark{1},
      {Y. Thomas Hou}\IEEEauthorrefmark{1}
}

    \IEEEauthorblockA{\IEEEauthorrefmark{1}Virginia Polytechnic Institute and State University, VA, USA}
    
    \IEEEauthorblockA{\IEEEauthorrefmark{2}Washington University in St. Louis, MO, USA}
    
    \thanks{An earlier version of this paper is accepted by 2020 IEEE International Conference on Computer Communications (INFOCOM 2020).}
    \thanks{\copyright 2020 IEEE. Personal use of this material is permitted. Permission from IEEE must be obtained for all other uses, including reprinting/republishing this material for advertising or promotional purposes, collecting new collected works for resale or redistribution to servers or lists, or reuse of any copyrighted component of this work in other works.}
}

\begin{document}



\maketitle
\begin{abstract}

Blockchain, the technology behind the popular Bitcoin, is considered a ``security by design" system as it is meant to create security among a group of distrustful parties yet without a central trusted authority. The security of blockchain relies on the premise of honest-majority, namely, the blockchain system is assumed to be secure as long as the majority of consensus voting power is honest. And in the case of proof-of-work (PoW) blockchain, adversaries cannot control more than 50\% of the network's gross computing power. However, this 50\% threshold is based on the analysis of computing power only, with implicit and idealistic assumptions on the network and node behavior. Recent researches have alluded that factors such as network connectivity, presence of blockchain forks, and mining strategy could undermine 
the consensus security assured by the honest-majority, but neither concrete analysis nor quantitative evaluation is provided. In this paper we fill the gap by proposing an analytical model to assess the impact of network connectivity on the consensus security of PoW blockchain under different adversary models. We apply our analytical model to two adversarial scenarios: 1) honest-but-potentially-colluding, 2) selfish mining. For each scenario, we quantify the communication capability of nodes involved in a fork race and estimate the adversary's mining revenue and its impact on security properties of the consensus protocol. Simulation results validated our analysis. Our modeling and analysis provide a paradigm for assessing the security impact of various factors in a distributed consensus system.
\end{abstract}

\section{Introduction}



Decentralization is a foundational principle for blockchain technology and distributed ledger system. 
Envisioned by Nakamoto, the pseudonymous creator of Bitcoin \cite{nakamoto2008bitcoin}, and later advocates,  
blockchain technology is secure-by-design and enables mutually distrustful parties to securely curate a shared blockchain through distributed consensus without relying on a central authority for bootstrapping the trust. Driven by incentives, consensus participants act in their self-interest to maximize rewards. Under such decentralized zero-trust setting, the security of distributed consensus relies on the premise of honest-majority, i.e. honest parties always control the majority of gross voting power in the consensus process, and in the case of proof-of-work (PoW) based blockchains, $50\%$ of gross computing (or ``mining'') power \cite{garay2015bitcoin}.
{\color{\rcolor}
This threshold is widely used for evaluating the risk of mining centralization in Bitcoin and Ethereum \cite{gervais2014bitcoin,beikverdi2015trend,srinivasan2017quantifying,gencer2018decentralization}.}

{\color{\rcolor}
However, the consensus security from honest-majority comes with two implicit assumptions.} First, all nodes have the same communication capability, i.e. propagating information throughout the network equally fast. Second, during a blockchain fork race, wherein several blocks of the same height compete for one place in the blockchain, all competitors have the equal chance of being the winner. 
In practice, the quality of connections often differ significantly across different network regions, as has been demonstrated by various measurements \cite{gencer2018decentralization,baumann2014exploring,miller2015discovering,neudecker2016timing}. Those residing in a highly connected cluster can disseminate blocks faster than those in a less connected region. This communication advantage translates into a higher chance of dominating a fork race, and has nontrivial consequence in the security of distributed consensus. As a result of this network advantage, the adversary will no longer require $50\%$ of gross mining power to undermine the consensus security. 

Following this observation, various blockchain scaling proposals and security analyses \cite{sompolinsky2013accelerating,sompolinsky2015secure,croman2016scaling} have identified the positive correlation between high blockchain fork rate and weak consensus security.
These works generally adopt the \textit{honest-but-potentially-colluding} threat model, in which any size of honest miners can join the collusion to compromise consensus security. Specifically, colluding miners seek to dominate the fork races with honest miners and achieve unfair mining gains. As a result, the colluding miners require less than $50\%$ of gross mining power to break the consensus. 
However, these security analyses are largely qualitative and do not look into the impact of the actual network connectivity or information propagation dynamics.


The security impact of information propagation dynamics in Bitcoin was studied quantitatively at the macro level in \cite{decker2013information}. It proposes a probabilistic model that estimates the average fork rate of the Bitcoin blockchain based on the measurement of how an average block propagates in the network. 
The authors then regard fork rate as a security measure of the blockchain network.
However, their model still assumes all miners have equal communication capability and equal chance of winning fork races, and does not consider the impact of heterogeneous network connectivity. It also does not provide a concrete case of how an adversary may exploit blockchain forks.

Another line of research focuses on adversarial strategies for selfish colluding parties \cite{eyal2018majority,sapirshtein2016optimal,gervais2016security,nayak2016stubborn}. In \emph{selfish mining}~\cite{eyal2018majority}, an adversary with superior communication capability can achieve unfair mining gains by strategically withholding and releasing blocks. It proactively creates blockchain forks that nullify the efforts of honest nodes. Although these works take into account the difference in fork winning chance between the adversary and honest miners, their analyses treat the adversary's communication capability as a preexisting parameter (denoted by $\gamma_{SM}$) rather than deriving it from the actual network connectivity profile. How the expansion process of selfish mining pool in the network affects its communication capability and overall consensus security is also an important issue but overlooked in the literature.

\begin{figure}
    \centering
    \includegraphics[width=3.2in]{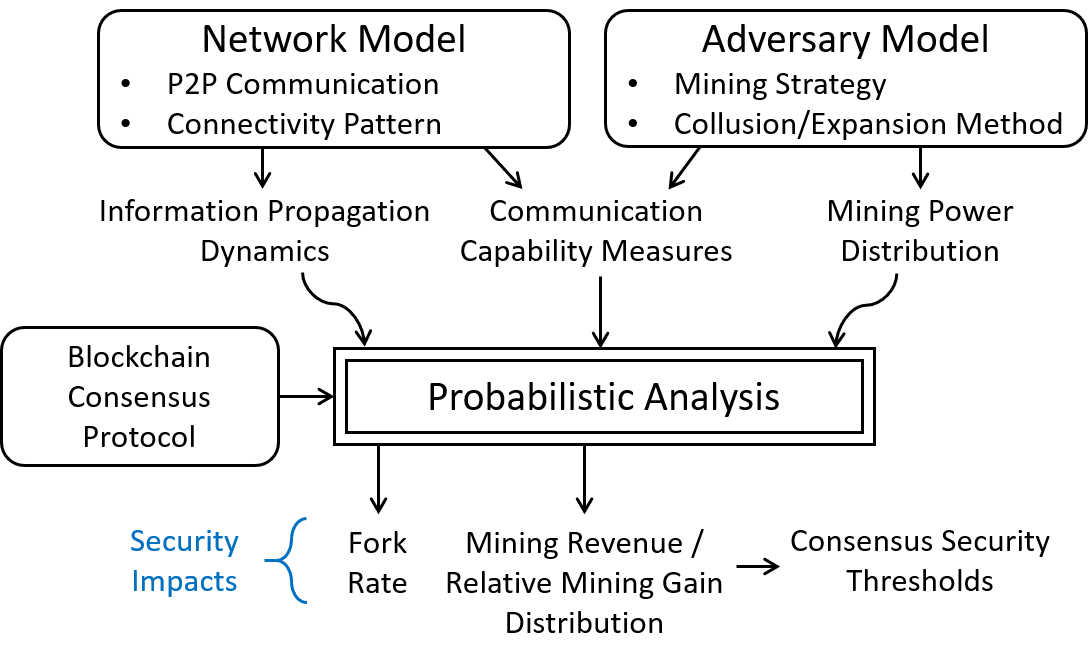}
    \caption{Proposed analytical model.}
    \label{fig:analysis-diagram}
\end{figure}

{\color{\rcolor}
Recognizing the lack of quantitative analysis on the security of distributed consensus from a network perspective,
in this paper we fill the gap by proposing a novel analytical model to assess the impact of network connectivity on the security of PoW-based blockchain consensus systems.}
An overview of our analytical model is given in Figure \ref{fig:analysis-diagram}.
The model captures network connectivity by a graph representation of the peer-to-peer network, and evaluates each node's communication capability from its network location and the adversary setting. The communication capability measures, combined with the consensus protocol specification and two other digests namely the information propagation dynamics and mining power distribution, are then used to quantitatively evaluate the security properties of the blockchain system.

The main contributions of this paper include:
\begin{itemize}
    \item We propose a novel analytical model to assess the impact of network connectivity on consensus security of PoW blockchain. In a nutshell, the model quantifies the communication capability of nodes involved in a fork race and derives the distribution of mining revenue, which is used for evaluating blockchain consensus security.
    \item We apply the analytical model to two adversarial scenarios, namely honest-but-potentially-colluding and selfish mining, and demonstrate that the lack of or excessively heterogeneous network connectivity leads to poor consensus security.
    \item We performed a thorough simulation experiment on PoW blockchain for each adversarial scenario. The simulation result matches the model prediction and validates our security analysis.
\end{itemize}


\section{Background}

\subsection{PoW Blockchain and Distributed Consensus}

In public blockchain systems exemplified by Bitcoin, all networked miner nodes (``nodes'' hereafter) work to curate a unified transaction history through distributed consensus. The transaction history is recorded in a chain of blocks in which every block contains a certain number of recently produced transactions. Every node seeks to generate the next block of the blockchain via a proof-of-work (PoW) process, namely, by finding an input to a cryptographic hash function that yields an output less than a target value. The input (i.e. the ``proof") is attached in the block header. New blocks are disseminated immediately to the network via peer-to-peer gossiping. All nodes reach consensus on only one block at each blockchain height according to the ``longest-chain rule'': choosing the chain with the highest valid block. Generation of the next block is aimed at prolonging the longest chain, 
{\color{\rcolor}
which shall always contain the most computational effort.} 
Theoretically, as long as the majority computing power is controlled by honest nodes,
{\color{\rcolor}
the distributed consensus achieves \emph{probabilistic finality} in that an accepted block could be discarded but with exponentially diminishing probability as the
blockchain grows \cite{bano2017consensus,xiao2019survey}.}
The above scheme is also known as Nakamoto Consensus, for its origin in the Bitcoin white paper \cite{nakamoto2008bitcoin}.

In practical blockchain network, consensus security is complicated by blockchain forks. Blockchain fork is a scenario that multiple blocks of the same height are propagating in the network simultaneously. Under the assumption that all nodes are honest and follow the consensus rules, blockchain fork is caused by block propagation delays in that node $j$ may generate a competing block before being aware of the existence of node $i$'s block of the same height. To resolve blockchain fork, the longest-chain rule dictates that whichever fork branch gets appended with a new block should be chosen; blocks in other branches are then discarded. In the presence of forks, the honest-majority premise can still ensure consensus security, under an assumption that all competing blocks in a fork have the equal chance of being the winner \cite{garay2015bitcoin}.

\subsection{Network Connectivity's Impact on Consensus Security}

Due to heterogeneous connectivity of the underlying peer-to-peer network, the equal-chance fork winning assumption does not hold true. A well-connected node, say node $i$, tends to have superior communication capability that allows it to disseminate information faster than a less-connected node, say node $j$. If node $i$ generates a new block, it takes a shorter time for node $i$ to propagate this block across the whole network and the rest of the network has a lower chance of generating a competing block. If node $j$ generates a competing block before node $i$'s block reaches $j$, node $i$'s communication advantage can still cause a larger share of the network to follow its block, which gives node $i$ a higher chance of winning the fork eventually. 
As a result, in the long run, well-connected nodes yield higher mining revenue than what is expected from their share of computing power. This discrepancy between the long-term mining revenue and the actual computing power of a node implies the possibility that a group of well-connected nodes with a minority fraction of computing power can harvest more than 50\% of gross mining revenue, which renders the security from the honest-majority premise vulnerable.


Besides exploiting naturally occurred forks, a well-connected adversary can achieve significantly higher mining gains by proactively creating forks. Selfish mining \cite{eyal2018majority} is one prominent example. Unlike an honest miner who publishes new block immediately after generation, a selfish mining attacker withholds newly generated blocks in a private chain, and strategically releases the private chain to the network whenever he sees his private chain's lead over the public chain decreases to a threshold.
The blockchain forks caused by the attacker's strategic private chain releases nullify the mining effort of honest nodes and create opportunities for the attacker to profit from its communication advantage.
The detailed selfish mining strategy and the communication advantage parameter $\gamma_{SM}$ will be introduced in Section \ref{sec:selfish-mine}.

\section{System Model}

\subsection{Network Model and Consensus Protocol}


We consider a peer-to-peer network of $N$ nodes represented by an undirected graph $\mathbf{G}=(\mathbf{V},\mathbf{E})$ and its adjacency matrix $\mathbf{A}$. $\mathbf{A}_{ij}=1$ indicates node $i$, $j$ share a peer relationship and can communicate in one hop. The PoW process and consensus based on the longest-chain-rule are modeled as follows.
To model the output randomness of the cryptographic hash function used for PoW, we assume each node $i$ generates new blocks according to Poisson process of rate $\pi_i$ per time unit $\delta$. Block generation of the whole network is characterized by the merged Poisson process of rate $\pi=\sum_i\pi_i$. Our model does not adjust mining difficulty. We consider a fixed set of consensus participants with fixed block generation rate.

Once some node $i$ generates $block_i(h)$ of blockchain height $h$, it disseminates $block_i(h)$ throughout the network via peer-to-peer gossiping. Other nodes decide on the acceptance of $block_i(h)$ according to the longest-chain rule. That is, if another node $k$ sees $block_i(h)$ while its local blockchain has already accepted $block_j(h)$ from node $j$ and $j\neq i$, it declares a fork at height $h$ and stops propagating $block_i(h)$. Conversely, if another node $l$ sees $block_i(h)$ before $block_j(h)$, it declares a fork at height $h$ and stops propagating $block_j(h)$. Once the two competing blocks completely stop propagating and the network partitions into two factions each of which advocates one block, we call this situation a \textit{fork stalemate}. And the two blocks are \textit{partially propagated}. A fork stalemate can be resolved by a new block of height $h+1$ subscribing either $block_i(h)$ or $block_j(h)$ and being \textit{fully propagated} in the network.

\begin{table}
    \centering
    \caption{Summary of Notations}
    \renewcommand{\arraystretch}{1.2}
    \begin{tabular}{|c|l|}
        
        \hline
        \multicolumn{2}{|c|}{Network and Model Parameters} \\ \hline
        $\mathbf{G}$ & The graph representation of the node network. \\ \hline
        $N$ & Number of nodes in $\mathbf{G}$. \\ \hline
        $\mathbf{A}$ & The adjacency matrix of $\mathbf{G}$. \\ \hline
        $\delta$ & Timeslot, also the time unit. \\ \hline
        $\pi$ & Block generation rate of the entire network ($\delta^{-1}$). \\ \hline
        $\pi_i$ & Block generation rate of node $i$ ($\delta^{-1}$). \\ \hline
        $\pi_{SM}$ & Block generation rate of the selfish mining pool ($\delta^{-1}$) \\ \hline
        \hline
        \multicolumn{2}{|c|}{Analyses} \\ \hline
        $\mathbf{U}_i(t)$ & Set of nodes unaware of node $i$'s block at time $t$ since its \\
        & generation. $|\mathbf{U}_i(t)|$ is the cardinality of $\mathbf{U}_i(t)$. \\ \hline
        $|\mathbf{U}_i(t)|_\pi$ & Combined block generation rate of nodes in $\mathbf{U}_i(t)$.  \\ \hline
        $P_{NC,i}(t)$ & Probability of the rest miners not proposing a competing \\
        & block by time $t$ of $i$'s block's propagation. \\ \hline
        $h(c)$ & Blockchain height of the $c^{th}$ canonization event \\ \hline
        $\tau_{ij}(t)$ & Minimum time for node $i$'s block to reach $j$ starting at \\
        & time $t$ from the generation of $i$'s block. \\ \hline
        $\omega_{i\succ j}(t)$ & Node $i$'s likelihood to win the fork race against node $j$ if \\
        & $j$ publishes a competing block at time $t$ from node $i$'s \\ 
        & block's generation. $\hat{\omega}_{i\succ j}(t)$ is an estimation. \\ \hline
        $\gamma_{SM}$ & Selfish mining pool's communication capability, i.e. the  \\
        & average fraction of honest mining power that will advocate \\
        & the pool's block after it releases private chain. \\ \hline
        $MR_i$ & Mining revenue of node $i$ as $\%$ of total canonized blocks. \\ \hline
        $RMG_i$ & Relative mining gain of node $i$. $RMG_i=\frac{MR_i-\pi_i/\pi}{\pi_i/\pi}$.  \\ \hline 
        \hline
        \multicolumn{2}{|c|}{Security Metrics} \\ \hline
        \mfr & Average fork rate of the whole network. \\ \hline
        \mat & $50\%$-attack threshold, i.e. minimum fraction of computing \\ & power controlled by adversarial nodes whose combined \mmr \\ 
        & exceeds $50\%$ of the total. \\ \hline
        \mpr & Profitability threshold, i.e. fraction of computing power  \\
        & controlled by the selfish mining pool when it first achieves \\
        & positive \mrm during its expansion. \\
        \hline
    \end{tabular}
    \renewcommand{\arraystretch}{1}
    \label{tab:notation}
\end{table}

\begin{figure}
    \centering
    \includegraphics[width=3.2in]{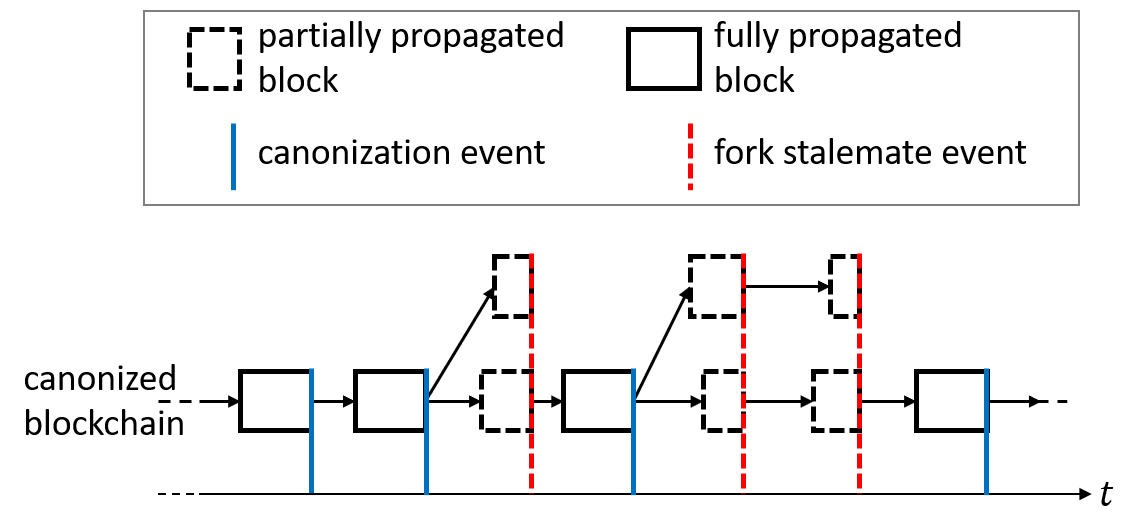}
    \caption{An example for blockchain canonization and fork stalemate events. Width of a block denotes its propagation period.}
    \label{fig:chain-canonization-stalemate}
    \vspace{-.2in}
\end{figure}










As for the finalization of blockchain, we consider the blockchain \textit{canonized} by height $h$ if a block of any origin $block_*(h)$ gets fully propagated in the network without encountering any competing block. We define the completion of $block_*(h)$'s propagation as a \textit{canonization event}. Essentially, a canonization event at height $h$ rejuvenates the PoW process as if the past forks and competitions never happened. Figure \ref{fig:chain-canonization-stalemate} shows an example for blockchain canonization and fork stalemate events. Note that the canonization concept is different from the probabilistic finality of Nakamoto Consensus, which considers consensus security a probabilistic measure. We will use canonization events as embedding points to estimate the mining revenue of each node in the Section \ref{sec:honest-mine}.


\subsection{Adversary Model}

\subsubsection{Honest-but-Potentially-Colluding} This adversarial scenario characterizes the practical case of the well-known $50\%$ attack. All nodes operate honestly by default, but the top miners can potentially collude so that their combined \textit{mining revenue (\mmr)} exceeds $50\%$ of the total. In our analysis, a well-connected node may obtain positive \textit{relative mining gain (\mrm)} and collude with other well-connected nodes. The mining revenue of a node can be viewed as its ``enhanced mining power'' in contrast to its raw computing power. In this scenario we are interested in the \textit{$50\%$-attack threshold (\mat)}, i.e. the minimum fraction of computing power controlled by adversarial nodes whose combined \mmr exceeds $50\%$ of the total.

\subsubsection{Selfish Mining} This adversarial scenario assumes there are a pool of nodes in the network performing the selfish mining strategy described in \cite{eyal2018majority}. We treat the selfish mining pool as a colluding consortium that expands among honest nodes. As the pool expands, it acquires the colluding nodes' computing power and external communication links. In this scenario, \mat denotes the fraction of computing power controlled by the pool when the pool first achieves $50\%$ of total \mmr during its expansion. We are also interested in the pool's \textit{profitability threshold (\mpr)}, which is the fraction of computing power controlled by the pool when it first achieves a positive \mrm.

\section{Analysis on Honest Mining}
\label{sec:honest-mine}

In this section we calculate the impact of network connectivity on blockchain fork rate and mining revenue distribution under the honest mining assumption. We then discuss the security provision under the honest-but-potentially-colluding adversarial scenario.

\subsection{Fork Rate}


Define $M_i$ as the event that node $i$ is the first to generate the next block at an arbitrary moment of no outstanding blockchain fork.
Denote the time for node $i$ to find a block by random variable $T_i$. Then $T_i\sim exponential(\pi_i)$ and 
\begin{equation}
    \label{eq:P_M_i}
    P(M_i) = P\{T_i<T_j,\forall j\neq i\} = \frac{\pi_i}{\pi}
\end{equation}
which can be conveniently derived from properties of Poisson processes. To facilitate the ensuing analysis, we consider the physical time slotted into basic time units of $\delta$.

Let the moment when event $M_i$ happens be time $0$. Denote $\mathbf{U}_i(t)$ the set of nodes unaware of node $i$'s block at time $t$, and $|\mathbf{U}_i(t)|_\pi$ the combined block generation rate of $\mathbf{U}_i(t)$. We have $|\mathbf{U}_i(0)|_\pi=\pi-\pi_i$ and $|\mathbf{U}_i(t)|_\pi=0$ when $t$ exceeds the minimum time needed for $i$'s block to reach all nodes. The probability that the rest of network does not generate a competing block by time $t$ can be written as:
\begin{equation}
    \label{eq:P_NC_i_t}
    P_{NC,i}(t) = \prod_{s=\delta}^{t} \big(1-|\mathbf{U}_i(s)|_\pi \big)
\end{equation}
Since $(1-|\mathbf{U}_i(s)|_M)_{s=\delta}^t$ is a positive increasing sequence bounded by $1$, thus $(P_{NC,i}(t))_{t=\delta}^\infty$ is a convergent sequence.

Then by the law of total probability, the average blockchain fork rate of the whole network is obtained by weighing $(1-\lim_{t\rightarrow\infty}P_{NC,i}(t))$ with $P(M_i),\forall i$:
\begin{equation}
    \label{eq:fork-rate}
    \begin{split}
        FR
        &= \sum_{i}P(M_i) \big(1-\lim_{t\rightarrow\infty}P_{NC,i}(t)\big) \\
        &= \sum_{i}\frac{\pi_i}{\pi}\Big(1 - \prod_{s=\delta}^{\infty} \big(1-|\mathbf{U}_i(s)|_\pi \big)\Big) \\
    \end{split}
\end{equation}


If $\pi\ll 1$, $N$ is large (e.g. $\pi$=$1/600$, $N$$\approx$$10,000$ in Bitcoin), mining power and network connectivity are evenly distributed, then $\forall i$ we have $\pi_i=\frac{\pi}{N}$, $|\mathbf{U}_i(s)|_\pi=\frac{\pi}{N}|\mathbf{U}_i(t)|=\pi\cdot \overline{ur}(t)$ wherein $\overline{ur}(t)$ denotes the average ratio of nodes uninformed of a new block at time $t$ since its generation.
Further assuming $\delta\rightarrow 0$, (\ref{eq:fork-rate}) reduces into the following form:
\begin{equation}
    \label{eq:fork-rate-simple}
    FR \approx 1-\Big(1-\pi \Big)^{\int_0^\infty \overline{ur}(t) dt}
\end{equation}
which is consistent with the average fork rate obtained by \cite{decker2013information}. The approximation $(1-ax)\approx(1-x)^a$ for small $x$ is used. 

\subsection{Mining Revenue and Relative Mining Gain}




Define a discrete-time random process $\{B_i(h)\}_{h=1,2,...}$ in which $B_i(h)=1$ if node $i$ is the block generator at height $h$ in the canonized blockchain; $0$ otherwise.
The mining revenue $MR_i$ and relative mining gain $RMG_i$ of node $i$ in the long term are defined as follows:
\begin{equation}
    \label{eq:MR_i}
    MR_i = \lim_{H\rightarrow\infty}\frac{1}{H}\sum_{h=1}^H B_i(h) 
\end{equation}
\begin{equation}
    RMG_i = \frac{MR_i - \pi_i/\pi}{\pi_i/\pi}
\end{equation}

Next we propose an estimation method for $MR_i$ via probabilistic analysis. 
Define another discrete-time random process $\{W_i(c)\}_{c=1,2,...}$, which is embedded right after each blockchain canonization event. Therefore there is no outstanding fork nor propagating block in the network when random variables $W_i(c)|_{c=1,2,...}$ are evaluated.  We further define $h(c)$ as the blockchain height of the $c^{th}$ canonization event and
\begin{equation}
    W_i(c) =
    \begin{cases}
    1 \hspace{.2in}\textit{if }B_i(h(c)+1)=1 \\
    0 \hspace{.2in}\textit{otherwise}
    \end{cases}
\end{equation}
Next we argue that the expectation of $W_i(c)$ at any epoch $c$, denoted $E[W_i]$, can be used to estimate $MR_i$ in a conservative manner.

\begin{proposition}
\label{prop:consersative-estimate}
\textit{
$W_i(c)|_{c=1,2,...}$ are independent and identically distributed (i.i.d.) and their common expectation $E[W_i]$ satisfies the following relation with $MR_i$:
\begin{equation}
    E[W_i]
    \begin{cases}
    \leq MR_i \hspace{.2in}\textit{if }E[W_i] \geq \frac{\pi_i}{\pi} \\
    > MR_i \hspace{.2in}\textit{otherwise}
    \end{cases}
\end{equation}
In other words, $E[W_i]-\frac{\pi_i}{\pi}$ is a conservative estimate of the mining gain/loss of node $i$. Moreover, the gap between $E[W_i]$ and $MR_i$ tightens as the overall fork rate \mfr decreases.
}
\end{proposition}
\begin{IEEEproof}[\textbf{A proof sketch}]
    Since $\{W_i(c)\}_{c=1,2,...}$ is embedded right after each blockchain canonization event when all previous forks are pruned and block propagation ceases, the competition for future blocks is oblivious of past block competitions. And block generation at each node is a memoryless process. Therefore, random variables $W_i(c)|_{c=1,2,...}$ are i.i.d and share a common expectation, denoted $E[W_i]$.
    
    For an arbitrary canonization interval $c\rightarrow c+1$, we consider the blocks within it: those of height $h(c)+1,...,h(c+1)$. First, $\frac{\pi_i}{\pi}$ evaluates the chance of $i$ being the first to generate a block. $E[W_i] > \frac{\pi_i}{\pi}$ implies that $i$ has a communication advantage over the network average which brings it positive mining gain. If $E[W_i] > \frac{\pi_i}{\pi}$ and $i$ wins block $h(c)+1$, it will continue with an enhanced communication advantage for the current fork race and have a higher chance of winning the subsequent blocks from $h(c)+2$ to $h(c+1)$. Conversely, if $E[W_i] < \frac{\pi_i}{\pi}$ and $i$ does not win block $h(c)+1$, the chance for $i$ to win any block from height $h(c)+2$ to $h(c+1)$ further decreases because of its aggravated communication disadvantage. Since $W_i(c)$ only considers the first block after canonization event $c$, using $E[W_i]$ to estimate $MR_i$ implies that $i$ would have an equal chance of winning any block from $h(c)+2$ to $h(c+1)$ just as winning $h(c)+1$. Therefore, $E[W_i]$ tends to underestimate (or overestimate) $MR_i$ if $E[W_i]>$ (or $<$) $\frac{\pi_i}{\pi}$.
    
    On the positive side, if the fork rate decreases, so is the number of blocks $h(c+1)-h(c)$ within the canonization interval $c\rightarrow c+1$. That is, there will be fewer blocks in a fork incident for $E[W_i]$ to under-/overestimate $MR_i$, and thus the former can achieve higher estimation accuracy.
\end{IEEEproof}

Next we derive $E[W_i]$. By the law of total expectation:
\begin{equation}
    \label{eq:winning-prob}
    E[W_i] = P(M_i)E[W_i|M_i] + \sum_{j\neq i} P(M_{j})E[W_i|M_j]
\end{equation}
We separated the summation because the two conditional events $W_i|M_i$ and $W_i|M_j$ occur under different conditions. 

$W_i|M_i$ consists of two subcases: 
\begin{itemize}
    \item \emph{No-fork win:} No conflicting blocks are proposed by the rest of the network during the propagation of node $i$'s block.
    \item \emph{Fork win:} Conflicting blocks are proposed by the rest of the network during the propagation of node $i$'s block, whereas node $i$'s block still wins.
\end{itemize}


The probability of no-fork win equals to $P_{NC,i}(\infty)$, as is evaluated by (\ref{eq:fork-rate}). The probability of fork win is slightly more complicated. During the propagation of node $i$'s block, the number of conflicting blocks generated by the rest of network at time slot $(t,t+\delta]$ conforms to a Bernoulli distribution with rate $|\mathbf{U}_i(t)|_M$. If node $j$ happens to generate a competing block during $(t,t+\delta]$, node $i$'s block will need to win the support of the majority computing power of the network before it encounters node $j$'s block in a stalemate. We denote the chance of node $i$ winning the fork under this condition by $\omega_{i\succ j}(t)$. Therefore:
\begin{equation}
    \label{eq:conditional-prob-1}
    \begin{split}
        &E[W_i|M_i] \\
        &= E[W_i,~\textit{no-fork win}|M_i] + E[W_i,~\textit{fork win}|M_i] \\
        &= \lim_{t\rightarrow\infty}P_{NC,i}(t) + \sum_{t=\delta}^{\infty}P_{NC,i}(t)\sum_{j\in\mathbf{U}_i(t)}\pi_j\omega_{i\succ j}(t) \\
    \end{split}
\end{equation}
Notably, in the derivation above we only considered two-prong forks for simplifying the analysis; likelihood of three or more-prong forks is negligible compared to that of two-prong forks.

In contrast, the conditional event $W_i|M_j$ in (\ref{eq:winning-prob}) can only happen via a fork race. That is, node $i$ needs to generate a competing block during the propagation of node $j$'s block, and eventually wins the fork race. Similarly to (\ref{eq:conditional-prob-1}), we have:
\begin{equation}
    \label{eq:conditional-prob-2}
    \begin{split}
        &E[W_i|M_j]
        = E[W_i,~\textit{fork win}|M_j] \\
        &= \sum_{t=\delta}^{\infty} P_{NC,j}(t)\pi_i\mathbb{1}_{\{i\in\mathbf{U}_j(t)\}}\big(1-\omega_{j\succ i}(t)\big) \\
    \end{split}
\end{equation}
$\mathbb{1}_{\{i\in\mathbf{U}_j(t)\}}$ is an indicator function, returning $1$ if the condition holds true; $0$ otherwise. The winning chance of node $i$ under this circumstance is $1-\omega_{j\succ i}(t)$. 




\textbf{Evaluating $\omega_{i\succ j}(t)$.~}
$\omega_{i\succ j}(t)$ essentially measures the communication advantage of node $i$ over $j$ when $j$ generates a competing block which starts the fork race. For $i$ to win the fork race against $j$, it has to have the majority of the network advocate its block before the two competing blocks end up in a stalemate. Let the moment when $i$ publishes its block be time $0$. Define $\tau_{ik}(t)$ as the minimum time for $i$'s block to propagate to node $k$ starting from time $t$. Then $\omega_{i\succ j}(t)$ can be estimated as follows:
\begin{equation}
    \label{eq:omega_t}
    \hat{\omega}_{i\succ j}(t) = \sum_{k}\frac{\pi_k}{\pi}\big(\mathbb{1}_{\{\tau_{ik}(t)<\tau_{jk}(0)\}} + \frac{1}{2}\mathbb{1}_{\{\tau_{ik}(t)=\tau_{jk}(0)\}}\big)
\end{equation}

\begin{figure}
    \centering
    \includegraphics[width=3in]{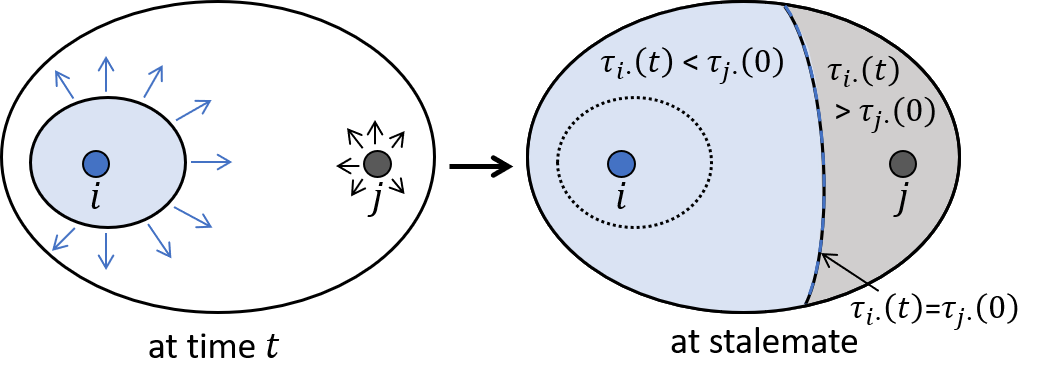} \vspace{-0.1in}
    \caption{Explanation of (\ref{eq:omega_t}). Light blue (grey) area denotes portion of the network that advocates $i$'s ($j$'s) block. $\hat{\omega}_{i\succ j}(t)$ is evaluated by the total computing power portion covered by light blue area at stalemate.}
    \label{fig:explain-omega}
\end{figure}

Figure \ref{fig:explain-omega} explains the calculation of $\hat{\omega}_{i\succ j}(t)$. As a result, we can finally obtain $E[W_i]$ by substituting (\ref{eq:P_M_i}), (\ref{eq:conditional-prob-1}), (\ref{eq:conditional-prob-2}), (\ref{eq:omega_t}) into (\ref{eq:winning-prob}).

\subsection{Security Analysis}
\label{subsec:honest-security-analysis}

We consider all nodes are honest-but-potentially-colluding. The fork rate \mfr provides an overall measure of how much mining power is wasted, while the $50\%$-attack threshold \mat measures the system's security in the worst case scenario that the colluding group consists of the highest mining revenue earners. Next we analyze how network connectivity impacts \mfr and \mat.

\subsubsection{Lower overall network connectivity leads to higher \mfr and lower \mat, thus weaker consensus security}
We assume the block generation rate $\pi_i$ is fixed for any node $i$. First, lower overall network connectivity means it takes longer for any node to disseminate a new block across the network. This can be caused by a protocol change that lowers the minimum peer number requirement. For the calculation in (\ref{eq:P_NC_i_t}) (\ref{eq:fork-rate}), this leads to a higher $|\mathbf{U}_i(s)|_\pi$, lower $\lim_{t\rightarrow\infty}P_{NC,i}(t), \forall i$, and thus higher \mfr. Moreover, a lower $\lim_{t\rightarrow\infty}P_{NC,i}(t)$ means that more of $MR_i$ is contributed by fork races and the distribution of mining revenue is deeper influenced by each node's communication capability. As a result, $MR_i$ moves farther from $\frac{\pi_i}{\pi}$ and \mat moves lower.

Notably, for a certain network connectivity profile, higher block generation rate across all nodes (thus a higher $\pi$) would lead to a higher $|\mathbf{U}_i(s)|_\pi,\forall i$ and have the same impact of lower overall network connectivity.

\subsubsection{Higher heterogeneity of network connectivity also leads to lower \mat}
We still assume the block generation rate $\pi_i,\forall i$ is fixed. Higher heterogeneity of network connectivity means there is a greater divergence of communication capability among nodes. For instance, if node $i$ resides in a highly-connected cluster in the network while node $j$ resides in a sparsely-connected region, $i$ will have a significant communication advantage over nodes in the sparse network region including $j$. As a result, $i$ can disseminate a block to majority of the network much faster than $j$. $\hat{\omega}_{i\succ j}(t)$, as is evaluated by (\ref{eq:omega_t}), will be close to $1$ and $\hat{\omega}_{j\succ i}(t)$ will be much lower than $0.5$. Therefore, $i$ can harvest more mining revenue from fork races than $j$ or other nodes in sparse network region. Consequently, $E[W_i]$ climbs higher above $\frac{\pi_i}{\pi}$ and $E[W_j]$ drops lower below $\frac{\pi_j}{\pi}$. This ultimately results in a more unequal \mmr distribution and hence lower \mat.
\section{Incorporating Network Connectivity into Selfish Mining Analysis}
\label{sec:selfish-mine}


In this section we evaluate the impact of network connectivity on selfish mining pool's communication capability and analyze its security implication under an expanding-consortium setting.

\subsection{Selfish Mining Strategy}

\begin{figure}
    \centering
    \includegraphics[width=3in]{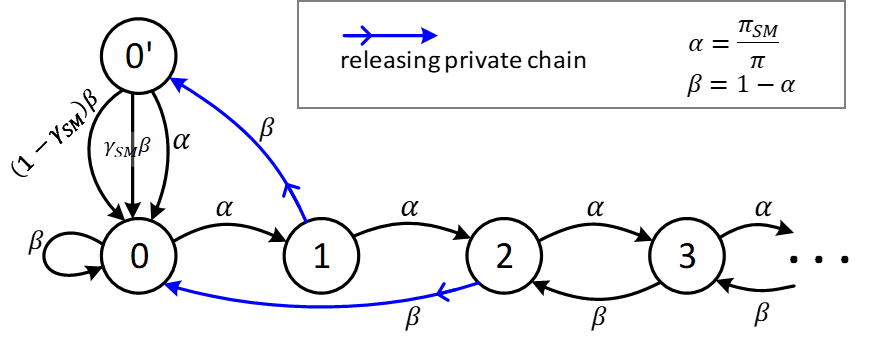}
    \caption{Selfish mining strategy in \cite{eyal2018majority}. State number denotes the private chain's lead over the honest chain. State transition is triggered by block generation.
    }
    \label{fig:selfish-mine}
\end{figure}

The core idea of selfish mining is to withhold newly generated blocks in a private chain, and release the private chain when the selfish mining pool sees the honest chain catch up close enough with the private chain. The detailed selfish mining strategy is illustrated in Figure \ref{fig:selfish-mine}, which we replicated from \cite{eyal2018majority} and added with more description. Let $\alpha$, $\beta$ be the computing power share of the selfish mining pool and the honest nodes. Then $\alpha=\pi_{SM}/\pi$ and $\beta=1-\alpha$, where $\pi_{SM}$ is the selfish mining pool's block generation rate. The state number denotes the private chain's lead over the honest chain. State transition is triggered by any block generation event. Transitions from state 1 to $0'$ and 2 to 0 are accompanied by the selfish mining pool's releasing the private chain. Any transition destined to state 0 marks a canonization event. $\gamma_{SM}$ is defined as the long-term average fraction of honest computing power that will advocate the selfish mining pool's private chain when the pool and an honest miner release competing blocks simultaneously.


To incorporate network connectivity into the analysis, we model the selfish mining pool's network function as follows:
\begin{itemize}
    \item Information exchanges within the selfish mining pool are instantaneous. The pool members are fully connected and synchronized. Any pool member who receives a new block from an honest node can make decision (changing state, switching chain, publishing the private chain) on behave of the entire pool.
    \item Once the selfish mining strategy determines to release the private chain, all pool members release the private chain to all peers simultaneously.
    \item Selfish mining pool members still relay blocks for honest miners, as long as the block does not trigger the pool to release its private chain. The reason is two-fold for the pool: to avoid suspicion of being a ``blackhole'' attacker, and to avoid network partitioning which would paralyze the blockchain system altogether. 
\end{itemize}

Based on this model, we consider $\gamma_{SM}$ the selfish mining pool's communication capability measure and evaluate it from the network connectivity profile. 




\subsection{Evaluating $\gamma_{SM}$ Using Betweenness Centrality}

Based on the assumption that nodes in the selfish mining pool are synchronous and can communicate with each other instantaneously, we treat these pool nodes as a fully-interconnected cluster and equivalently a super node denoted by \smp, which preserves the pool members' all external communication links to the remaining honest nodes. 
We show that a betweenness centrality measure of \smp within the network accurately evaluates $\gamma_{SM}$.



\begin{proposition}
\label{prop:gamma-betweenness}
\textit{
$\gamma_{SM}$ can be evaluated by the mining power weighted betweenness centrality measure of \smp:
\begin{equation}
    \label{eq:sm-gamma}
    \begin{split}
        &\gamma_{SM}=\\
        &\sum_{i\neq\smp} \frac{\pi_i}{\pi-\pi_{SM}} \sum_{j\neq i \neq \smp} \frac{\pi_j}{\pi-\pi_{SM}-\pi_i} \frac{\sigma(i,j|\smp)}{\sigma(i,j)}   
    \end{split}
\end{equation}
wherein $\sigma(i,j)$ is the number of shortest paths between $i$ and $j$, and $\sigma(i,j|\smp)$ is the number of such paths that pass through \smp.
}
\end{proposition}
\begin{IEEEproof}[\textbf{A proof sketch}]
Let $i$ and $j$ denote a pair of honest nodes, with $i$ being the miner of a new block which triggers \smp to release its private chain according to the selfish-mine strategy. For $j$ to switch to \smp's private chain instead of accepting $i$'s new block, the highest block of \smp's private chain must be propagated to $j$ before $i$'s new block. In the graph model, this is necessitated by \smp residing on a shortest communication path between $i$ and $j$. Therefore, $\frac{\sigma(i,j|\smp)}{\sigma(i,j)}$ gives the likelihood that \smp delivers its private chain to $j$ ahead of $i$'s block. The weight $\frac{\pi_i}{\pi-\pi_{SM}} \frac{\pi_j}{\pi-\pi_{SM}-\pi_i}$ evaluates the pair $(i,j)$'s mining power contribution to $\gamma_{SM}$ among all pairs of honest nodes. As a result, the mining power weighted betweenness centrality measure of \smp computes the average fraction of honest mining power that will advocate \smp's block after \smp releases its private chain, thus accurately evaluates $\gamma_{SM}$.
\end{IEEEproof}



Equation (\ref{eq:sm-gamma}) can be conveniently computed with the Brandes algorithm \cite{brandes2001faster}.
If there are $M$ honest nodes and they have equal mining power, i.e. $\pi_i=\frac{\pi-\pi_{SM}}{M},\forall i\neq SM$, then the weight $\frac{\pi_i}{\pi-\pi_{SM}} \cdot \frac{\pi_j}{\pi-\pi_{SM}-\pi_i}$ becomes $\frac{1}{M(M-1)}$ and (\ref{eq:sm-gamma}) reduces to the standard normalized betweenness centrality measure. 

With $\gamma_{SM}$ obtained, the calculation of the selfish mining pool's mining revenue follows the procedure of \cite{eyal2018majority}. Notably, the mining revenue of pool is proportional to $\gamma_{SM}$.

\subsection{Security Analysis}
\label{subsec:sm-security-analysis}

We consider the selfish mining pool as an expanding consortium among the network of honest nodes. Under this setting, we discuss how network connectivity affects $\gamma_{SM}$ and consensus security w.r.t. security thresholds \mat and \mpr.


\subsubsection{Lower overall network connectivity leads to higher $\gamma_{SM}$, lower \mat, and lower \mpr, thus less secure against selfish mining}
Lower overall network connectivity leads to reduced communication capability of both the selfish mining pool and honest nodes. However, since the selfish mining pool consists of originally honest nodes and preserve all their external communication links, communication capability reduction of an average honest node will be more significant than that of \smp. Therefore, \smp will be residing in the shortest communication paths of more honest pairs if the overall network connectivity lowers, yielding a higher $\gamma_{SM}$ for every $\alpha$ value. Consequently this yields lower \mat and \mpr.

\subsubsection{Compared to \mat, \mpr is more sensitive to network connectivity changes}
According to \cite{eyal2018majority}, \mpr is reached much earlier than \mat for any $\gamma_{SM}$.
Due to the gradualism of selfish mining pool's expansion, the rate of the selfish mining pool harvesting new external communication links initially increases, then gradually slows down as the pool takes in more nodes. Thus as $\alpha$ grows from $0$ to $50\%$, $\gamma_{SM}$ grows quickly at first then slow down as it comes closer to 1. Also the mining revenue of selfish mining pool is proportional to $\gamma_{SM}$. Therefore, a moderate reduction of overall network connectivity would lead to significant decrease in \mpr, but limited decrease in \mat.
\mpr is also a more practical security measure than \mat in the sense that once the pool size hits \mpr, joining the pool will be financially attractive to the remaining honest nodes in the network.

\subsubsection{Selfish mining pool can take advantage of heterogeneity of network connectivity to achieve lower AT50 and PRTH}
If the selfish mining pool is aware of the peer-to-peer network's topology, it can prioritize its expansion into well-connected regions of the network to maximize the growth of $\gamma_{SM}$ and its mining revenue. As a result, heterogeneity of network connectivity can be exploited by selfish mining pool to achieve lower \mat and \mpr.


\section{Evaluation}

We conducted simulation experiments to validate our model and security analysis. The simulation program was written in Python and follows a time-driven fashion and takes the following as input: graph representation of the network, block generation rates of all nodes, adversarial setting (honest or selfish mining), and simulation time (slots).



\begin{figure}
    \centering
    \subfigure[$G_R(1000,4)$]{
        \centering
        \includegraphics[width=1.1in]{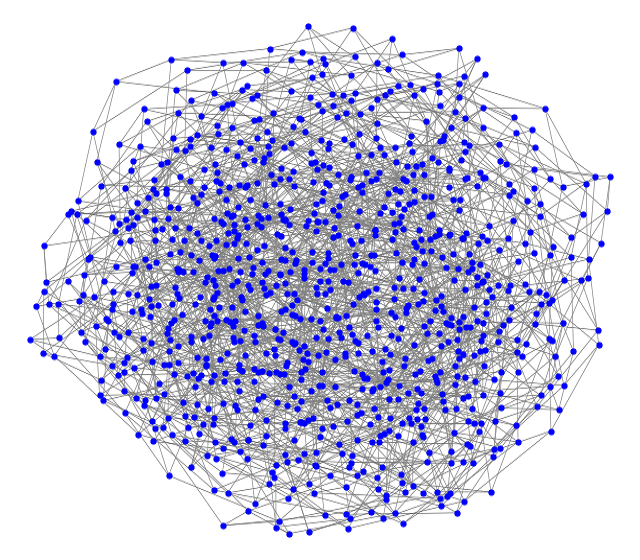}
        \label{fig:graph-visual-a}
    }
    \hspace{-0.18in}
    \subfigure[$G_R(1000,4)_{F10}$]{
        \centering
        \includegraphics[width=1.1in]{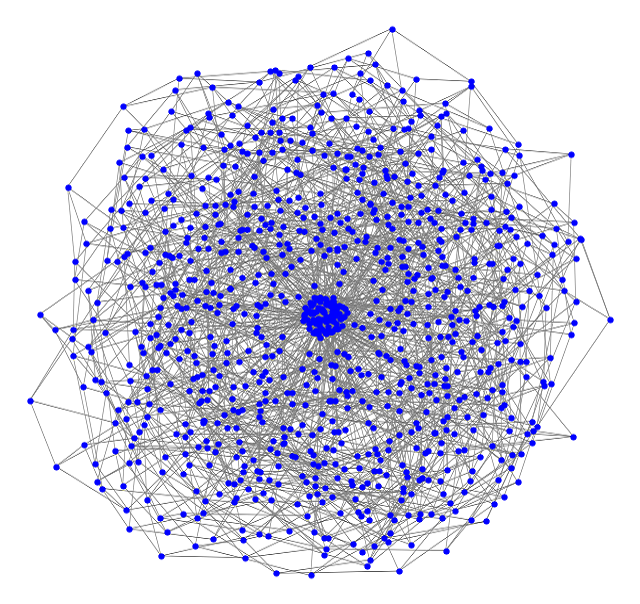}
        \label{fig:graph-visual-b}
    }
    \hspace{-0.2in}
    \subfigure[$G_E(1000,4)$]{
        \centering
        \includegraphics[width=1.1in]{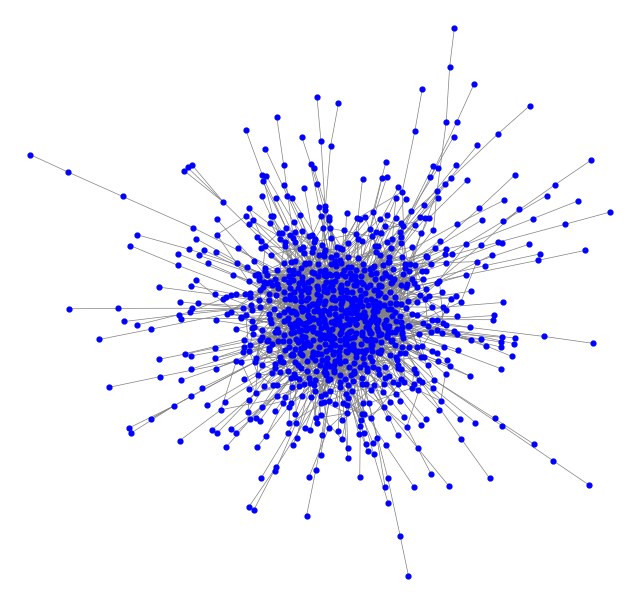}
        \label{fig:graph-visual-c}
    }
    \subfigure[$G_R(100,4)$]{
        \centering
        \includegraphics[width=1.1in]{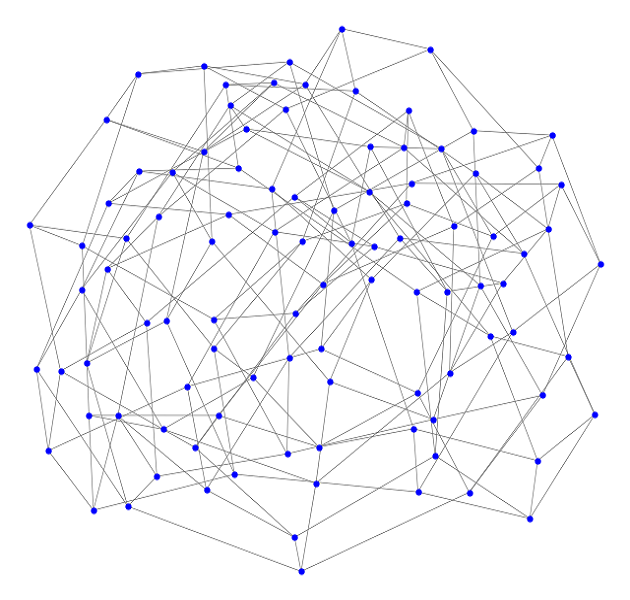}
        \label{fig:graph-visual-d}
    }
    \hspace{-0.18in}
    \subfigure[$G_R(1000,8)_{F10}$]{
        \centering
        \includegraphics[width=1.1in]{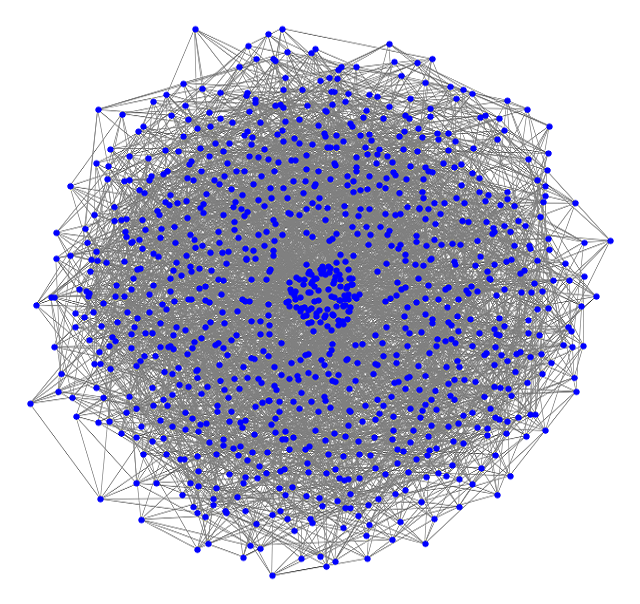}
        \label{fig:graph-visual-e}
    }
    \hspace{-0.2in}
    \subfigure[$G_E(1000,8)$]{
        \centering
        \includegraphics[width=1.1in]{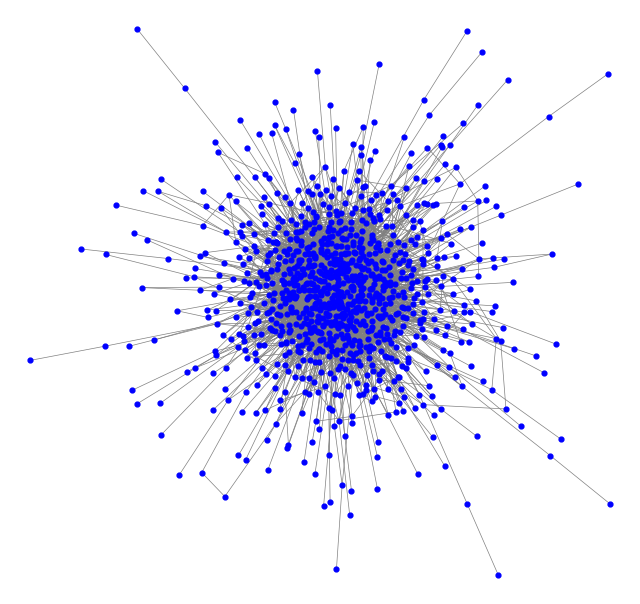}
        \label{fig:graph-visual-f}
    }
    \\ 
    \caption{Visualization of six network graphs used in our experiments. Dot represents mining node, grey line segment represents communication link.
    }
    \label{fig:graph-visual}
    \vspace{-10pt}
\end{figure}

\begin{figure*}
    \centering
    \subfigure[$G_E(1000,4),\pi=0.1$]{
        \centering
        \includegraphics[width=1.74in]{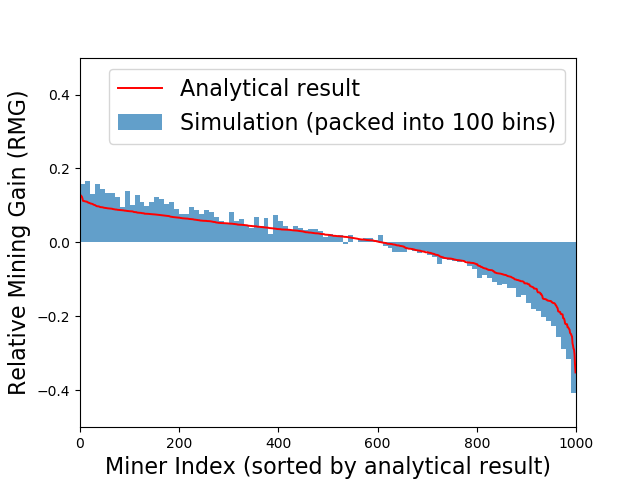}
        \label{fig:RMG-exp-a}
    }
    \hspace{-0.2in}
    \subfigure[$G_E(1000,4),\pi=0.05$]{
        \centering
        \includegraphics[width=1.74in]{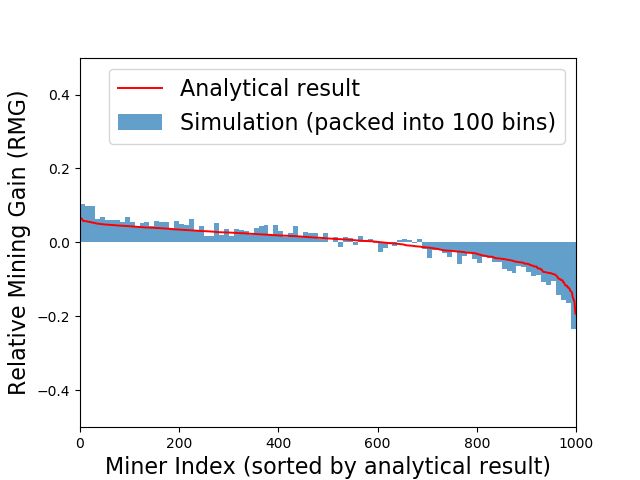}
        \label{fig:RMG-exp-b}
    }
    \hspace{-0.2in}
    \subfigure[$G_E(1000,8),\pi=0.1$]{
        \centering
        \includegraphics[width=1.74in]{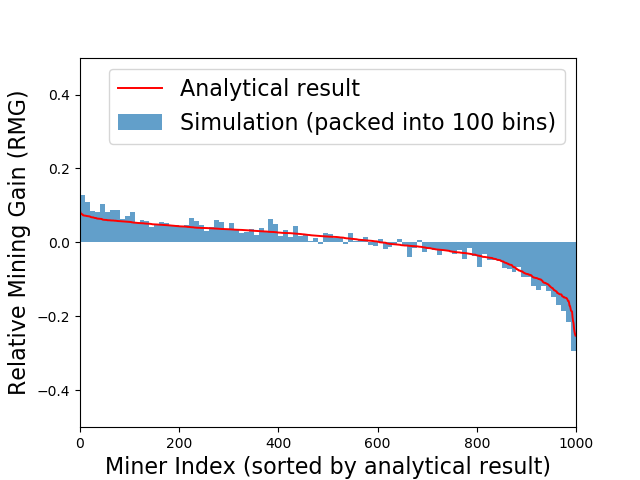}
        \label{fig:RMG-exp-c}
    }
    \hspace{-0.2in}
    \subfigure[$G_E(1000,8),\pi=0.05$]{
        \centering
        \includegraphics[width=1.74in]{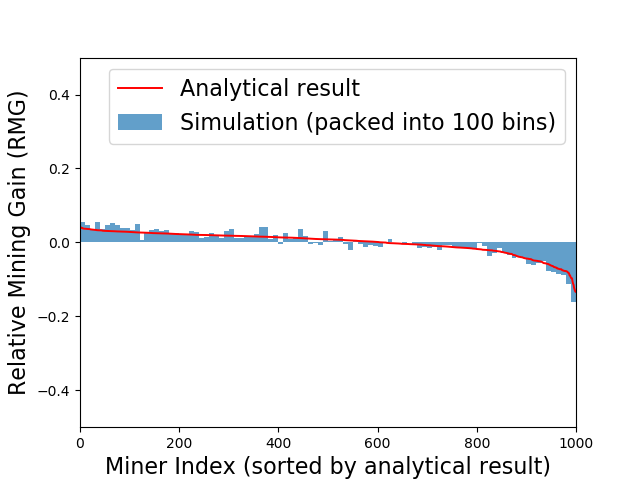}
        \label{fig:RMG-exp-d}
    }\\ 
    \vspace{-0.1in}
    \subfigure[$G_R(1000,4)_{F10},\pi=0.1$]{
        \centering
        \includegraphics[width=1.74in]{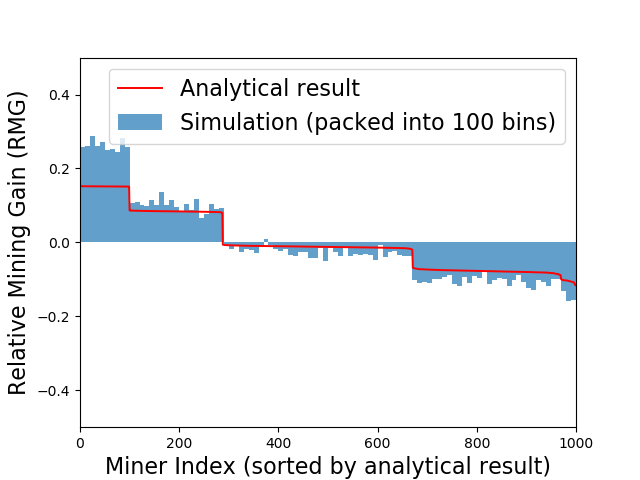}
        \label{fig:RMG-exp-e}
    }
    \hspace{-0.2in}
    \subfigure[$G_R(1000,4)_{F10},\pi=0.05$]{
        \centering
        \includegraphics[width=1.74in]{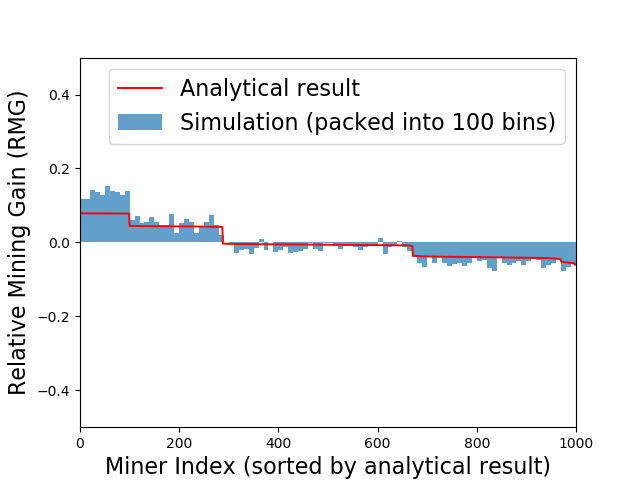}
        \label{fig:RMG-exp-f}
    }
    \hspace{-0.2in}
    \subfigure[$G_R(1000,8)_{F10},\pi=0.1$]{
        \centering
        \includegraphics[width=1.74in]{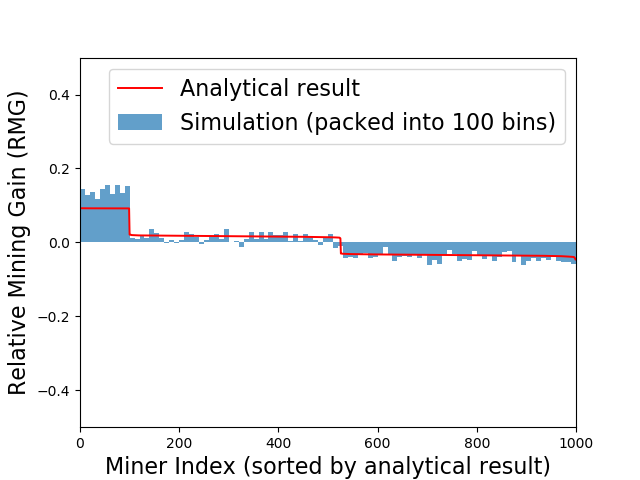}
        \label{fig:RMG-exp-g}
    }
    \hspace{-0.2in}
    \subfigure[$G_R(1000,8)_{F10},\pi=0.05$]{
        \centering
        \includegraphics[width=1.74in]{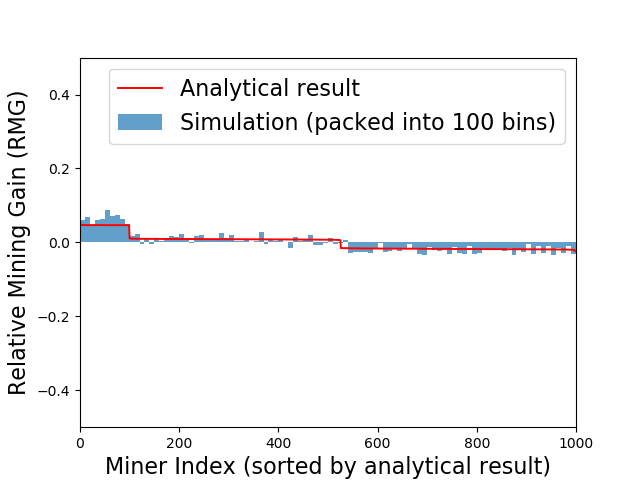}
        \label{fig:RMG-exp-h}
    }\\ 
    \caption{Relative mining gain (\mrm) results of eight experiments. }
    \label{fig:RMG-exp}
\end{figure*}

\begin{table*}
    \centering
    \footnotesize
    \caption{Honest Mining Experiment Result Corresponding to Figure \ref{fig:RMG-exp}}
    \renewcommand{\arraystretch}{1.1}
    \begin{tabular}{c|cc|cc|cc|c}
        \toprule
         & \multicolumn{2}{c|}{\textbf{Configuration}} & \multicolumn{5}{c}{\textbf{Metrics}} \\
         & Graph$(N,D)$ & $\pi$ & FR-SIM & FR-ANA  & AT50-SIM & AT50-ANA  & RMSE \\
        \midrule
        a & $G_E(1000,4)$ & $0.1$ & $0.3148$ & $0.3136$ &  $459/1000$ & $470/1000$ & $0.0325$ \\
        
        b & $G_E(1000,4)$ & $0.05$ & $0.1773$ & $0.1670$ & $478/1000$ & $484/1000$ & $0.0205$ \\
        
        c & $G_E(1000,8)$ & $0.1$ & $0.2409$ & $0.2248$ & $475/1000$ & $479/1000$ & $0.0187$ \\
        
        d & $G_E(1000,8)$ & $0.05$ & $0.1315$ & $0.1159$ & $487/1000$ & $489/1000$ & $0.0123$ \\
        
        e & $G_R(1000,4)_{F10}$ & $0.1$ & $0.3117$ & $0.3099$ & $457/1000$ & $470/1000$ & $0.0423$ \\
        
        f & $G_R(1000,4)_{F10}$ & $0.05$ & $0.1758$ & $0.1649$ & $479/1000$ & $485/1000$ & $0.0232$ \\
        
        g & $G_R(1000,8)_{F10}$ & $0.1$ & $0.2309$ & $0.2124$ & $480/1000$ & $484/1000$ & $0.0195$ \\
        
        h & $G_R(1000,8)_{F10}$ & $0.05$ & $0.1250$ & $0.1090$ & $490/1000$ & $491/1000$ & $0.0113$ \\
        \bottomrule
    \end{tabular}
    \renewcommand{\arraystretch}{1}
    \label{tab:RMG-exp-comparison}
    \vspace{-8pt}
\end{table*}

\subsection{Setup}

We use three types of network graph for evaluation with the following notations: 
\begin{itemize}
    \item $G_R(N,D)$: a regular graph with $N$ nodes and degree $D$.
    \item $G_R(N,D)_{FX}$: a $G_R(N,D)$ but with the first $X\%$ of nodes being fully-interconnected.
    \item $G_E(N,D)$: a graph with $N$ nodes and exponentially distributed node degrees with an average $D$.
\end{itemize}
The latter two graph types are designed to simulate different heterogeneous network connectivity profiles. The six network graphs used in our experiments are visualized in Figure \ref{fig:graph-visual} with Python package NetworkX \cite{hagberg2008exploring}.
To evaluate the impact of network connectivity and provide a fair ground for comparing security thresholds, we assign all nodes the same block generation rate: $\pi_i=\frac{\pi}{N}, \forall i$ while using $\pi$ as a system input.

The following metrics are used for evaluation:
\mfr, \mat and \mpr as security metrics,
\textit{rooted mean square error (RMSE)} between analytical \mrm distribution and simulated \mrm distribution as the model accuracy metric.

\subsection{Honest Mining Experiment}

We performed eight experiments on four network graphs with different settings. Each experiment was run for 10 million timeslots. The configuration and evaluation results are shown in Table \ref{tab:RMG-exp-comparison} and the \mrm histograms are shown in Figure \ref{fig:RMG-exp}. We made the following observations:




\textit{1) The analytical \mrm result conservatively estimates the simulation result. The accuracy improves when $\pi$ decreases or $D$ increases.~}
For instance, the obvious gap between analytical result and simulation in Figure \ref{fig:RMG-exp-e} demonstrates the fully-interconnected top-$10\%$ nodes have a significantly higher block winning chance than that expected by $E[W_i]$. Nonetheless, as is shown in Table \ref{tab:RMG-exp-comparison}, for either graph type when $\pi$ decreases from $0.1$ to $0.05$ or $D$ increases from $4$ to $8$, the fork rate decreases and so does \textit{RMSE}. This validates Proposition \ref{prop:consersative-estimate} that the estimation accuracy improves when fork rate drops.

\textit{2) FR decreases and AT50 increases when $\pi$ decreases or $D$ increases.~}
This validates the security analysis in \ref{subsec:honest-security-analysis} that higher overall network connectivity or lower block generation rate leads to stronger consensus security in the presence of potentially colluding nodes.

\subsection{Selfish Mining Experiment}

\begin{figure*}
    \centering
    \hspace{0.16in}
    \subfigure[Analytical $\gamma_{SM}$. 
    ]
    {
        \centering
        \includegraphics[width=2.80in]{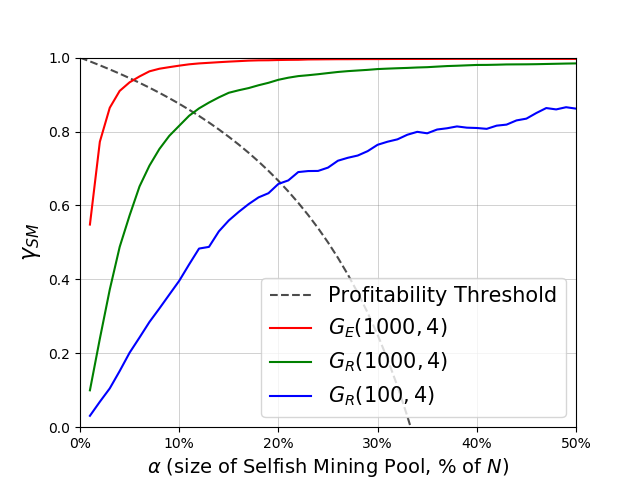}
        \label{fig:gamma_sm_vs_alpha}
    }
    \subfigure[Analytical $RMG$ and simulation.]
    {
        \centering
        \includegraphics[width=2.81in]{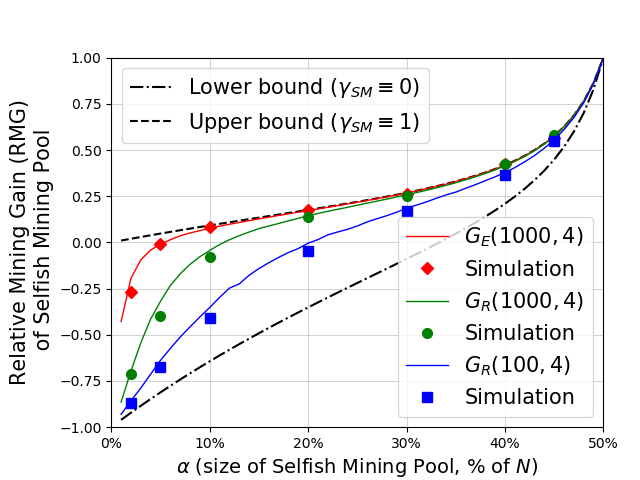}
        \label{fig:rmg_sm_vs_alpha}
    }
    \caption{Comparison of simulation and analytical result for selfish mining.}
    \label{fig:sm-experiments}
\end{figure*}

We switched the adversary setting from honest to selfish mining and performed three experiments. Each experiment targeted a certain network graph and consisted of seven simulations, each took an $\alpha$ value from $\{2,5,10,20,30,40,45\}\%$ and ran for 10 million timeslots. Figure \ref{fig:gamma_sm_vs_alpha} shows the analytical result of $\gamma_{SM}$. The configuration and evaluation results are shown in Table \ref{tab:SM-exp-comparison} and Figure \ref{fig:rmg_sm_vs_alpha}. For graph $G_E(1000,4)$ we configured the selfish mining pool to expand from the highest-degree node to lower-degree nodes in a descending order. This was purposed as an example of selfish mining pool's expansion strategy. \textit{RMSE} here measures the averaged estimation accuracy of the analytical \mrm over all the seven $\alpha$ values.






\begin{table*}
    \centering
    \caption{Selfish Mining Experiment Result Corresponding to Figure \ref{fig:sm-experiments}}
    \renewcommand{\arraystretch}{1.1}
    \begin{tabular}{c|cc|cc|cc|c}
        \toprule
         & \multicolumn{2}{c|}{\textbf{Configuration}} & \multicolumn{5}{c}{\textbf{Metrics}} \\
         & Graph $(N,D)$ & $\pi$ & PRTH-SIM & PRTH-ANA  & AT50-SIM & AT50-ANA & RMSE  \\
        \midrule
        1 & $G_E(1000,4)$ & $0.01$ & $52/1000$ & $55/1000$ &  $369/1000$ & $369/1000$ & $0.0290$ \\
        
        2 & $G_R(1000,4)$ & $0.01$ & $122/1000$ & $114/1000$ & $368/1000$ & $370/1000$ & $0.0338$ \\
        
        3 & $G_R(100,4)$ & $0.01$ & $22/100$ & $21/100$ & $38/100$ & $38/100$ & $0.0311$ \\
        \bottomrule
    \end{tabular}
    \renewcommand{\arraystretch}{1}
    \label{tab:SM-exp-comparison}
    \vspace{-8pt}
\end{table*}

To estimate the thresholds \mat and \mpr, for each of the three experiments we fitted the simulated \mrm points using degree-7 polynomials and obtained \mat and \mpr using the fitted curve. The following observations are made:

\textit{1) The analytical result matches simulation.~}
The close match between analytical \mrm results and simulation results in Figure \ref{fig:rmg_sm_vs_alpha} validates our betweenness centrality-based calculation of $\gamma_{SM}$. 

\textit{2) When $N$ decreases from $1000$ to $100$, PRTH changes more dramatically than AT50.~}
This validates our security analysis that \mpr is more sensitive to network connectivity changes. Particularly, the analytical curves of $\gamma_{SM}$ in Figure \ref{fig:gamma_sm_vs_alpha} demonstrates that as the selfish mining pool size expands from $\alpha=0$ to $\alpha=50\%$, $\gamma_{SM}$ grows quickly at first then gradually slows down when it comes closer to 1.

\textit{3) As is shown in Figure \ref{fig:gamma_sm_vs_alpha}, $\gamma_{SM}$ rapidly crosses above the \mpr curve and grows close to 1 when the selfish mining pool expands in $G_E(1000,4)$.~}
This demonstrates the feasibility for selfish mining pool to choose a particular expansion strategy to exploit the heterogeneity of network connectivity for faster revenue growth.

\section{Discussion and Future Work}

\textit{On Potential Model Deficiency:~}
In our model we only consider two-prong forks. That is, at most two blocks of the same height could be propagating in the network concurrently. Though the possibility of three-prong forks or more is significantly lower than two-prong forks, it could still affect the long-term mining revenue distribution, especially when the network is large and block propagation delays are high.
To tackle this we would need a more delicate model that accounts the progress of all competing branches in a fork. 

\textit{On Model Practicality:~}
In practical blockchain networks, it can be challenging to monitor the block propagation progress (i.e. $U_i(t)$). To overcome this difficulty, a block propagation progress-agnostic model is needed to estimate the communication capability and forecast the revenue of a node via congregate network statistics. Furthermore, the permissionless network is structurally volatile and may conform to a scale-free expansion pattern \cite{baumann2014exploring}. It is possible to model structural changes of the network with a certain random process and evaluate its impact on information propagation and consensus security.
The impact of a selfish mining pool's internal communication routine is also a potential issue to explore.
\section{Conclusion}
\label{sec:conclude}

We presented a modeling study on the impact of network connectivity on consensus security of PoW blockchain under two adversarial scenarios, namely honest-but-potentially-colluding and selfish mining. For the first scenario, we demonstrated the communication advantage of a node over its competitors in a fork race and provided a method to estimate its long-term mining revenue and relative mining gain. For the second scenario, we introduced a practical model for the selfish mining pool's network functions and showed that the communication capability of selfish mining pool, $\gamma_{SM}$, can be accurately evaluated by the mining power-weighted betweenness centrality measure. For both scenarios, we showed that low network connectivity and excessive heterogeneity of network connectivity lead to poor consensus security.
Our modeling and analysis can serve as a paradigm for assessing the security impact of various factors in a distributed consensus system.
In future work we will incorporate more realistic network settings into our model and extend the analysis to other blockchain systems.



\section*{Acknowledgment}
This work was supported in part by US National Science Foundation under
grants CNS-1916902 and CNS-1916926.



\bibliography{reference}
\bibliographystyle{IEEEtran}


\end{document}